\documentstyle{AGILEpsr}

\input epsf.sty
\input epsfig.sty
\input psfig.sty

\def \AAP #1 #2 {{\em Astron. Astrophys.\/} {\bf #1}, #2}
\def \AAL #1 #2 {{\em Astron. Astrophys. Lett.\/} {\bf #1}, L#2}
\def \AAR #1 #2 {{\em Astron. Astrophys. Rev.\/} {\bf #1}, #2}
\def \AAS #1 #2 {{\em Astron. Astrophys. Suppl. Ser.\/} {\bf #1}, #2}
\def \AJ #1 #2 {{\em Astron. J.\/} {\bf #1}, #2}
\def \ANNREV #1 #2 {{\em Ann. Rev. Astron. Astrophys.\/} {\bf #1}, #2}
\def \APJ #1 #2 {{\em Astrophys. J.\/} {\bf #1}, #2}
\def \APJL #1 #2 {{\em Astrophys. J. Lett.\/} {\bf #1}, L#2}
\def \APJS #1 #2 {{\em Astrophys. J. Suppl.\/} {\bf #1}, #2}
\def \APSS #1 #2 {{\em Astrophys. Space Sci.\/} {\bf #1}, #2}
\def \ASR #1 #2 {{\em Adv. Space Res.\/} {\bf #1}, #2}
\def \BAIC #1 #2 {{\em Bull. Astron. Inst. Czechosl.\/} {\bf #1}, #2}
\def \JSQRT #1 #2 {{\em J. Quant. Spectrosc. Radiat. Transfer\/} {\bf #1}, #2}
\def \MN #1 #2 {{\em Mon. Not. R. Astr. Soc.\/} {\bf #1}, #2}
\def \MEM #1 #2 {{\em Mem. R. Astr. Soc.\/} {\bf #1}, #2}
\def \PLR #1 #2 {{\em Phys. Lett. Rev.\/} {\bf #1}, #2}
\def \PASJ #1 #2 {{\em Publ. Astron. Soc. Japan\/} {\bf #1}, #2}
\def \PASP #1 #2 {{\em Publ. Astr. Soc. Pacific\/} {\bf #1}, #2}
\def \NAT #1 #2 {{\em Nature\/} {\bf #1}, #2}
\def \SAIT #1 #2 {{\em Mem.\ Soc.\ Astron.\ It.\/} {\bf #1}, #2}
\def \MESS #1 #2 {{\em The Messenger\/} {\bf #1}, #2}
\def \ASTRNACH #1 #2 {{\em Astron. Nach.\/} {\bf #1}, #2}
\def \AGPSR #1 #2 {{\em ASI Special Publication\/} {\bf #1}, #2}
\newcommand\arcsec{\mbox{$^{\prime\prime}$}}%
\newcommand{\gtap}{\mathrel{\hbox{\rlap{\lower.55ex \hbox {$\sim$}}
                   \kern-.3em \raise.4ex \hbox{$>$}}}}
\newcommand{\ltap}{\mathrel{\hbox{\rlap{\lower.55ex \hbox {$\sim$}}
                   \kern-.3em \raise.4ex \hbox{$<$}}}}

\begin{opening}

\title{X-ray and $\gamma$-ray observations of millisecond pulsars }
\author{L. Kuiper$^{1}$, W. Hermsen$^{1}$}
\institute{$^1$SRON National Institute for Space Research, Utrecht, The Netherlands}
\date{} 
\end{opening}

\begin{document}

\oddpagefooter{}{}{} 
\evenpagefooter{}{}{} 
\medskip  

\begin{abstract}
The launch of several sensitive X-ray and $\gamma$-ray instruments during the last
decade heralded a new era in the research of millisecond pulsars. The current number
of millisecond pulsars detected in the X-ray spectral window is about 30, including
those located in globular clusters, which represents a significant fraction of
the total number of spin-down powered pulsars emitting high-energy radiation.
In this paper the observational X/$\gamma$-ray status is reported for a subset of X-ray
emitting millisecond pulsars which show high-energy tails: PSR B1821-24, PSR J0218+4232
and PSR B1937+21. The prospects for future detection of these 3 millisecond pulsars at 
soft (INTEGRAL) and hard (AGILE/GLAST) $\gamma$-rays are discussed.    
\end{abstract}

\medskip

\section{Introduction}

\noindent
Millisecond pulsars are rapidly spinning neutron stars with (very) small period derivatives
(P $\ltap$ 15 ms; $\dot{\hbox{\rm P}} \ltap 10^{-17}$ s/s). Pulsars belonging to the 
millisecond-pulsar (MSP) group are preferentially located in binary systems (75\%), commonly 
with a white dwarf companion. It is generally believed that these sources represent the 
final stage of low-mass X-ray binary evolution.
The link between the X-ray binary and MSP stage is formed by the so-called 
accretion-driven X-ray MSPs of which there have been detected six members
nowadays (Wijnands 2003). 
\begin{figure}[h]
 \hbox{\hspace{0.0cm}\psfig{figure=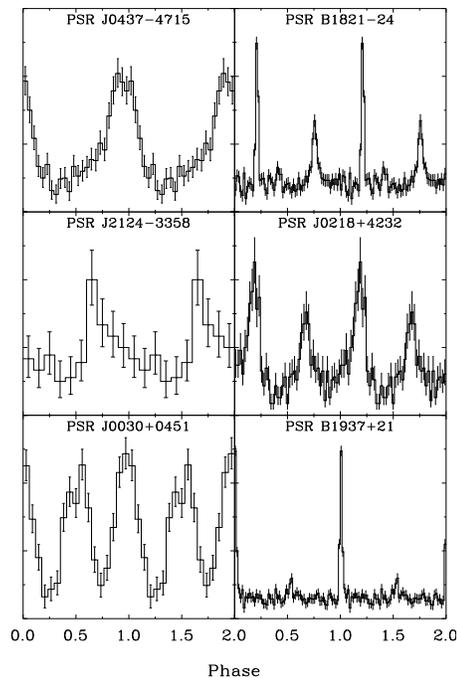,bbllx=125pt,bblly=135pt,bburx=465pt,bbury=665pt,clip=,height=9cm,width=6cm,angle=0}
        \hspace{0.1cm}{\parbox{65mm}{\vspace{-9cm}
        \caption{X-ray pulse profiles for the six spin-down powered MSPs
showing X-ray modulation at the rotational period. The pulsars in the left panels belong to
the Class-I X-ray emitting MSPs characterized by low X-ray luminosities, 
soft spectra and broad pulses. In contrast, the right panels show the Class-II MSPs with 
high X-ray luminosities, hard spectra and narrow pulses.\hfill
           \label{profile_collage}}}}}
\end{figure}
The first X-rays from a spin-down powered MSP had been detected by Becker 
\& Tr\"umper (1993) for PSR J0437-4715. With this detection a new spectral window, besides 
the radio band, was opened in which MSPs can be studied.
Currently, there are 12 plus 16 (the latter in globular cluster $^{47}$Tuc; Grindlay et al. 2003) X-ray
emitting spin-down powered MSPs detected. Of these, six (probably one more; PSR J1012+5307)
are known to emit {\em pulsed} X-ray emission. The X-ray pulse profiles for this small group
composed of PSR J0437-4715, PSR J2124-3358, PSR J0030+0451, PSR B1821-24, PSR J0218+4232 and
PSR B1937+21 are shown in Fig. \ref{profile_collage}. This small group can on observational X-ray
characteristics be subdivided in two classes, Class-I and II (Kuiper et al. 2000). 

The Class-I members - PSR J0437-4715, PSR J2124-3358 and PSR J0030+0451 - have low X-ray luminosities
(1-10 keV; $< 10^{30}$ erg/s), soft predominantly thermal X-ray spectra (e.g. Halpern et al. (1996) , Zavlin et al. (2002), Sakurai et al. (2001), 
Becker et al. (2002)) and broad pulses. It is generally believed that for this class the (modulated) 
X-ray emission originates 
from the polar cap region which is heated by backwards flowing particle currents in the 
pulsar's magnetosphere (thus thermal emission). Note, that in view of the high age of millisecond 
pulsars in general X-ray emission due to global surface cooling forms not a viable mechanism.

In contrast, the Class-II members - PSR B1821-24, PSR J0218+4232 and PSR B1937+21 - have high
X-ray luminosities (1-10 keV; $> 10^{32}$ erg/s), hard power-law shaped X-ray spectra and emit narrow X-ray pulses. These characteristics
point to a non-thermal origin related to physical processes taking place in the compact 
magnetosphere of a MSP.
Because the latter class shows hard X-ray spectral tails up to $\sim 20$ keV and because for 
one member, PSR J0218+4232, pulsed high-energy $\gamma$-rays have been detected (Kuiper et al. 
(2000,2002)) the MSPs in this class are promising candidates to be detected 
at high-energy $\gamma$-rays ($> 100$ MeV) by the future $\gamma$-ray missions AGILE and GLAST
and at soft $\gamma$-rays (15-1000 keV) by INTEGRAL, launched October 2002. 
Therefore, in the remainder of this review we will focus on the Class-II MSPs.

\section{The Class-II X-ray millisecond pulsars}


\subsection{PSR B1821-24}

The first MSP discovered in a globular cluster was PSR B1821-24 in M28/NGC 6626 (Lyne et al. 1987). 
It is a solitary pulsar with a pulse period of $\sim 3.05$ ms 
and a period derivative of $1.6\times 10^{-18}$ s/s. Its characteristic age of $3\times 10^7$ year
and spin-down luminosity of $2.2\times 10^{36}$ erg/s make it the ``youngest" and most energetic
MSP known today. Pulsed X-ray emission from this MSP was discovered
by Saito et al. (1997) using ASCA GIS data (0.7-10 keV). Two narrow pulses with different intensities
and separated $\sim 0.55$ in phase (phase distance between main X-ray pulse and secondary X-ray 
pulse) were visible in the pulse phase distribution. Also at soft X-rays (0.1-2.4 keV) the 
X-ray pulsations were detected (ROSAT HRI; Danner et al. 1997). The hard nature of the
pulsed emission suggested a magnetospheric origin. 

Inaccuracies in the ASCA and ROSAT clocks prevented an ``absolute" alignment of the X-ray and 
radio profiles. A short 6.5 ks observation performed on 16 September 1996 with the PCA aboard 
RXTE made such a comparison in absolute phase possible (Rots et al. 1998): the X-ray (2-16 keV) 
and radio profiles were nearly aligned, the main X-ray pulse XP1 lags the radio pulse by $\sim 60\mu$s.
It is interesting to note that the XP1 location is coincident with the phase location, where radio giant pulses
occur (Romani \& Johnston 2001).
Subsequent long exposure RXTE observations (85.5 ks during 10 -- 12 February 1997; 32.8 ks during 
12 -- 13 November 1999) of PSR B1821-24 made it possible to study the X-ray timing and spectral 
properties in detail (e.g. Kawai et al. 1999; Kuiper et al. 2004). The resulting high statistics 
X-ray pulse profile (2-20 keV) is shown in the upper right panel of Fig. 1. The phase separation 
is $0.546(2)$ and the FWHM's of the X-ray pulses are 0.018(1) and 0.046(6) for pulse 1 and 2, 
respectively, fitting two symmetric Lorentzians and a flat background to the X-ray pulse profile. 
This corresponds to very small pulse widths of $55\mu$s and $140\mu$s for pulse 1 and 2, respectively!
\begin{figure}[t]
  \centerline{
    \hbox{\psfig{figure=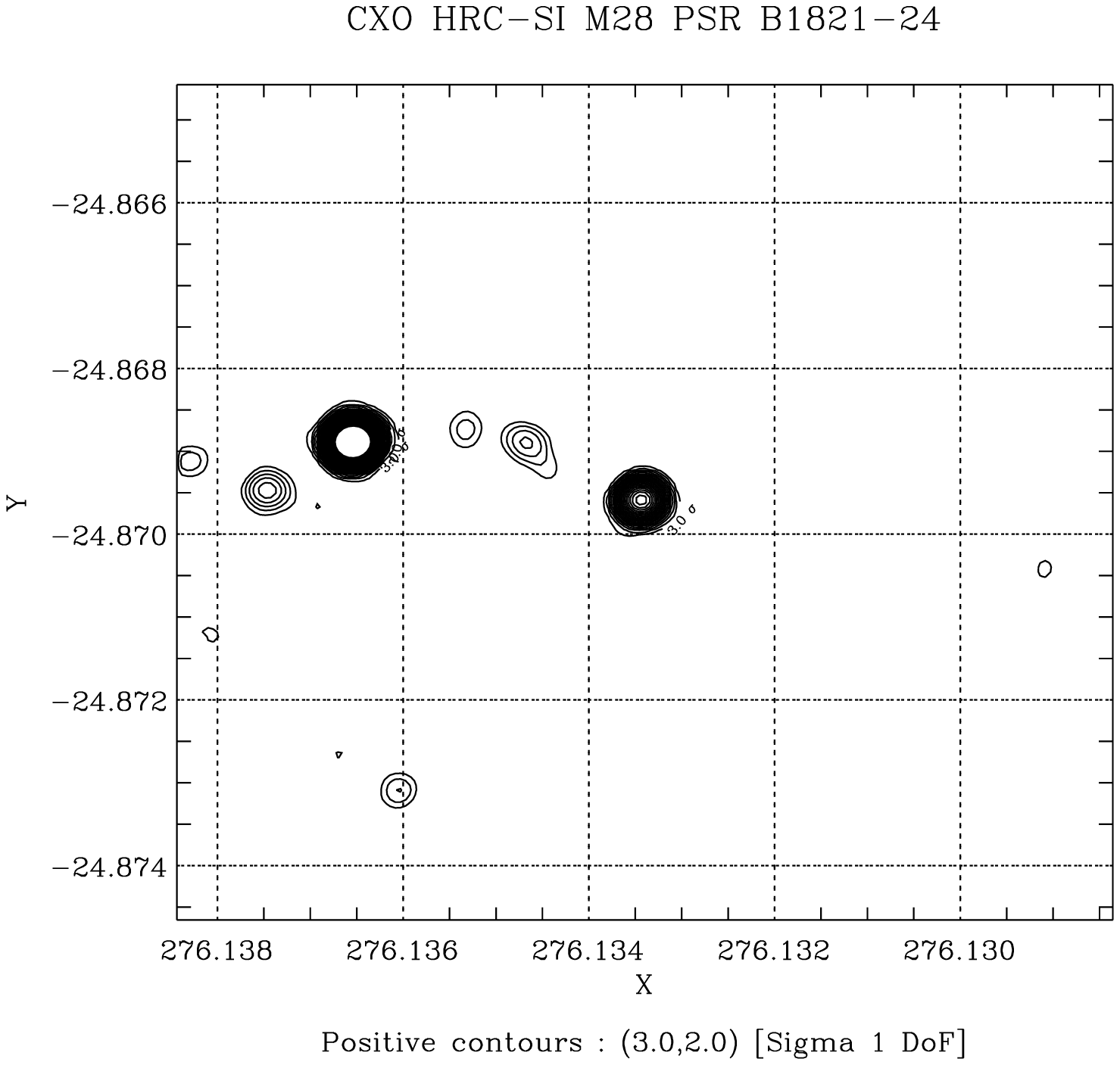,bbllx=34pt,bblly=358pt,bburx=460pt,bbury=720pt,clip=,height=6.6cm,width=6.5cm,angle=0}
          \hspace{0.25cm}
          \psfig{figure=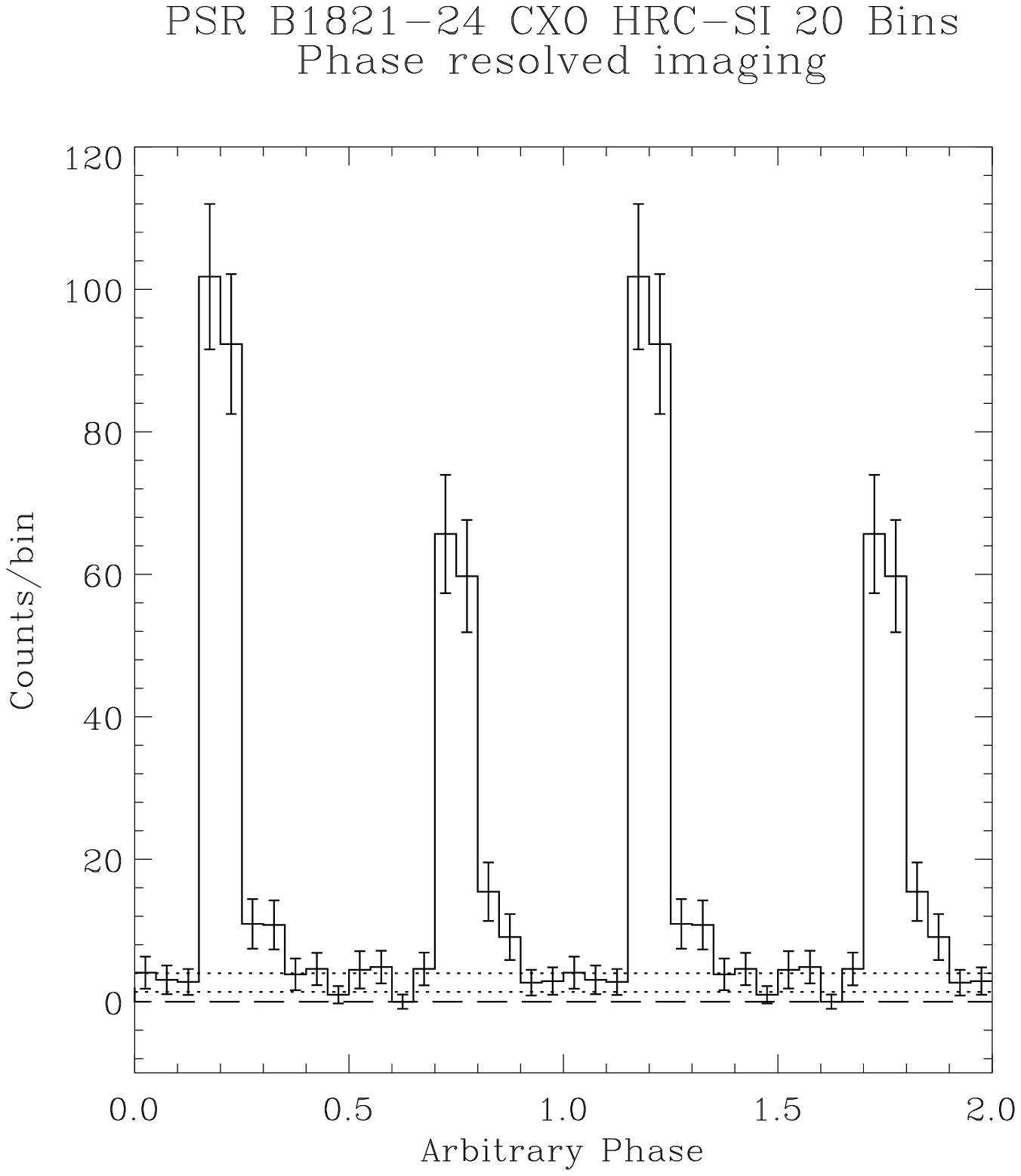,bbllx=110pt,bblly=217pt,bburx=500pt,bbury=615pt,clip=,height=6.45cm,width=6.5cm,angle=0}
         }
             }
  \caption[h]{(left) ML-image of the crowded core of M28 using HRC-SI data (0.08-10 keV). The fully resolved central point 
                     source is PSR B1821-24. The X and Y axes represent the RA-TAN and DEC-TAN for epoch 2000 in degrees.
                     (right) Pulse profile of PSR B1821-24 obtained from a phase-resolved ML-imaging analysis adopting 20 phase slices.
                     The dotted lines represent the $\pm 1 \sigma$ uncertainty boundaries for the DC-emission (12\%$\pm$6\%). }
\end{figure}
Combining the data from the three RXTE observations and two ASCA observations a joint spectral analysis (0.8-20 keV) 
of the pulsed spectrum yielded a photon index of 1.13(2) assuming an absorbing column N$_{\hbox{\scriptsize \rm H}}$ of 
$1.6\times 10^{21}$cm$^{-2}$ (Kuiper et al. 2004).

The location of PSR B1821-24 near the crowded core of globular cluster M28 requires high-resolution
X-ray imaging to derive e.g the spectrum and pulsed fraction {\em not} polluted by contributions from neighbouring sources. 
Such high-resolution imaging observations have recently been performed with the Chandra ACIS-S and HRC-SI instruments.
The total (pulsed plus DC) 0.5-8 keV spectrum obtained with the ACIS-S could be described
by a power-law with photon index $1.20\pm_{0.13}^{0.15}$ (Becker et al. 2003). Both the photon index and
normalization of the total spectrum are consistent with the corresponding numbers for the pulsed component
alone, leaving little room for a DC-component. 

A 50.8 ks Chandra HRC-SI observation of PSR B1821-24 on 8 November 2002 made it possible to resolve the source and to determine
the pulsed fraction of PSR B1821-24 for the integral 0.08-10 keV energy band. A value of
$0.85\pm0.03$ was obtained (Rutledge et al. 2004). We derived an HRC-SI image from this observation using
a Maximum Likelihood (ML) imaging method (Kuiper et al. 1998a) demonstrating the power of high-resolution imaging
in crowded fields (see Fig. 2 (left panel)). The pulsed fraction (0.08-10 keV) obtained from a phase-resolved
ML-imaging analysis is 0.88(7), consistent with the value obtained by Rutledge et al. (2004) using a different
approach. The pulse profile resulting from the phase-resolved imaging study adopting 20 phase slices of equal width 
is shown in Fig. 2 (right panel). The $\pm 1\sigma$ uncertainty boundaries on the DC-emission are shown by dotted
lines. 

Finally, at soft and high-energy $\gamma$-rays only upper limits have been reported (CGRO OSSE: Schroeder et al. (1995)
; CGRO EGRET: Fierro et al. (1995)).


\begin{figure}[h]
 \hbox{\hspace{0.0cm}\psfig{figure=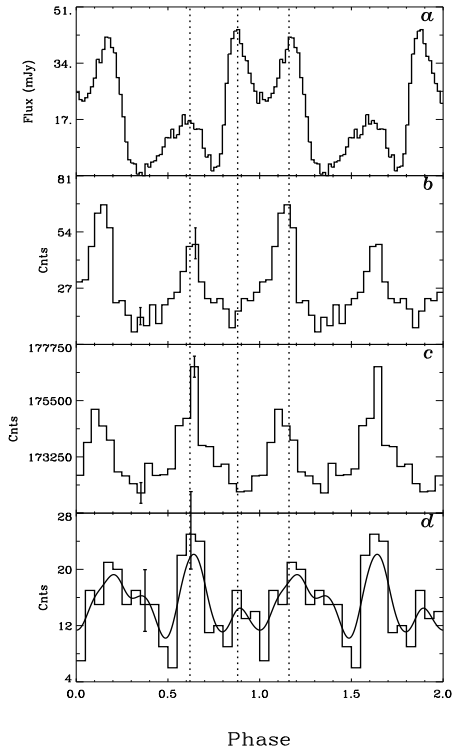,bbllx=140pt,bblly=148pt,bburx=420pt,bbury=655pt,clip=,height=10cm,width=6cm,angle=0}
        \hspace{0.1cm}{\parbox{65mm}{\vspace{-9.5cm}
        \caption{Multi-wavelength pulse profiles of PSR J0218+4232 in absolute phase. From top to bottom:
  Radio 610 MHz, X-rays: Chandra HRC-SI 0.08-10 keV and RXTE PCA 2-16 keV, and high-energy $\gamma$-rays:
  CGRO EGRET 100-1000 MeV. Note the morphology change of the pulse profile at X-rays (panels b and c).
  The three dotted lines indicate the locations of the radio pulses, while the
  smooth curve superposed on the pulse phase distribution in the bottom panel represents the Kernel Density
  estimator (see Kuiper et al. 2002).\hfill
           \label{candi_he_profiles}}}}}
\end{figure}

\subsection{PSR J0218+4232}

PSR J0218+4232 has been discovered by Navarro et al. (1995). It is a luminous radio pulsar
with a pulse period of 2.3 ms orbiting a degenerate companion in $\sim 2.0$ days. Its radio
profile is broad and complex with three pulses (Navarro et al. 1995; Stairs et al. 1999) and
never goes off (DC radio component). 

Soft X-ray emission (0.1-2.4 keV) from this source was detected by Verbunt et al. (1996) in
a short ROSAT HRI observation performed in August 1995. Its pulsed nature was discovered by
Kuiper et al. (1998b) in a 98 ks ROSAT HRI follow-up observation (July 1997). The 0.1-2.4 keV
pulse profile showed two peaks separated in phase by 0.47(1). The pulsed fraction (0.1-2.4 keV) 
was 37$\pm$13\% indicating a large DC (non-pulsed) contribution.
Spectral (phase-resolved) information at X-rays has been obtained by the MECS instrument (1.6-10 keV)
aboard BeppoSAX. The major findings (Mineo et al. 2000) from this $\sim 83$ ks observation are:
a) the total pulsed emission is (very) hard (photon-index $0.61\pm0.32$); b) the morphology of
the X-ray profile changes as a function of energy; c) the DC-fraction is a function of energy
and above $\sim 4$ keV it is consistent with being zero.

At high-energy $\gamma$-rays Kuiper et al. (2000) reported the likely detection of pulsed
$\gamma$-ray emission from this pulsar using CGRO EGRET data. The source was seen in the map below 1 GeV and 
the two pulses visible in the 100-1000 MeV pulse profile (deviation from uniformity $\sim 3.5\sigma$) turn out to be aligned
with two of the three radio pulses.

At X-rays the absolute timing accuracy for both the ROSAT HRI and BeppoSAX MECS did not
allow to phase relate the radio and X-ray profiles. Two $\sim 75$ ks Chandra observations,
one with the HRC-I and a second with the HRC-SI, made it possible to determine the phase alignment 
of the X-ray and radio profiles and to study the spatial extent of the X-ray 
emission from PSR J0218+4232 (Kuiper et al. 2002). The X-ray emission appeared to be point-like i.e. there were no indications
for extended emission beyond $1\arcsec$ scales (diameter). The two non-thermal X-ray pulses in the 0.08-10 keV
profile (see Fig. 1 middle right panel) are coincident with two of the three radio pulses {\em and}
with the two $\gamma$-ray pulses, increasing the detection significance of the pulsed $\gamma$-ray
signal to $\sim 4.9\sigma$. The pulsed fraction in the integral 0.08-10 keV band is $0.64\pm0.06$.
Combining the DC informations from the ROSAT HRI, BeppoSAX MECS and Chandra HRC-SI yielded, assuming
a power-law shape model, a photon index in the range 1.3 -- 1.85 for the non-pulsed emission.
This is indeed considerably softer than the spectrum of the pulsed component.
 
RXTE observed PSR J0218+4232 for $\sim 200$ ks between 26-12-2001 and 7-1-2002. Pulsed X-ray emission
was detected up to $\sim 20$ keV in both the RXTE PCA and RXTE HEXTE data (Kuiper et al. 2004). 
The X-ray pulse alignment relative to the radio profile verified the Chandra alignment results 
(see Fig. 3 for a multi-wavelength comparison of the pulse profile of PSR J0218+4232). A spectral analysis of the
total pulsed signal combining (revisited) BeppoSAX MECS and RXTE PCA data yielded a photon-index of
$0.99\pm0.03$ over the entire MECS/PCA energy range, one of the hardest spectra measured so far for any of
the spin-down powered pulsars detected at X-rays! 

Finally, it is interesting to note that recently also for this pulsar large amplitude pulses have been detected
at radio frequencies (Joshi et al. 2004). 


\begin{figure}[t]
  \centerline{\psfig{figure=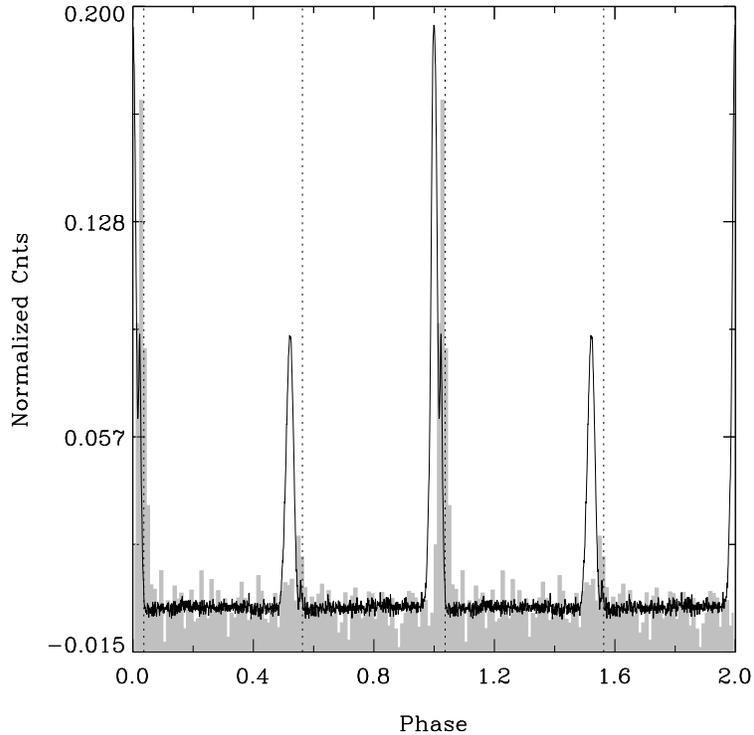,bbllx=80pt,bblly=205pt,bburx=500pt,bbury=625pt,clip=,height=10cm,width=10cm,angle=0}}
  \caption{RXTE PCA 2-10 keV pulse profile (histogram; light grey) with superposed a 1410 MHz
           radio profile obtained from Effelsberg observations in absolute phase. The X-ray main and
           secondary pulses lag the radio main and interpulse by $\sim 44\mu$s and $\sim 51\mu$s,
           respectively. The X-ray pulse locations are coincident with the locations where the radio
           giant pulses are found (dotted lines).\label{psrb1937_radio_xray}}
\end{figure}
\subsection{PSR B1937+21}

PSR B1937+21 was the very first MSP discovered at radio frequencies (Backer et al. 1982). With its
spin period of 1.56 ms this solitary MSP is still the most rapidly rotating neutron star currently known. A remarkable
feature is the existence of giant radio pulses (Wolszczan et al. 1984; Sallmen \& Backer 1995; Cognard et al. 1996) 
similar to those of the Crab pulsar.
\noindent
An initial attempt to search for X-ray emission from this MSP was performed by Verbunt et al. (1996).
They found no soft X-ray emission from the pulsar position in a $\sim 24$ ks ROSAT HRI observation.
However, at harder X-rays ($> 2$ keV) Takahashi et al. (2001) found eventually X-ray pulsations 
in ASCA GIS data. One very narrow pulse was visible in the X-ray pulse profile with an indication
for a weaker second pulse. This second weaker pulse was convincingly detected in a 78.5 ks BeppoSAX
MECS observation (Nicastro et al. 2002, 2004) performed in 2001. 
Spectral analyses of the total and pulsed signals (1.3-10 keV; MECS) showed a significant difference in the hardness
of the spectra: total emission power-law index $1.94\pm_{0.11}^{0.13}$; pulsed emission $1.21\pm_{0.13}^{0.15}$.
This is evidence for the presence of an underlying soft DC-component whose spatial extension is compatible with the 
MECS point spread function, and its contribution to the total emission amounts $\sim 45\%$ in the 1.3-4 keV band (Nicastro et al. 2004)

In February 2002 PSR B1937+21 was observed by RXTE for about 140 ks. In the high-resolution PCA 2-10 keV pulse profile 
two (very) sharp pulses were visible separated 0.526(2) in phase (Cusumano et al. 2003). The X-ray main and 
secondary pulses lag the radio main and interpulse by just $\sim 44\mu$s and $\sim 51\mu$s, respectively (see Fig. 4). 
Their locations are coincident with the locations where at radio frequencies giant radio pulses occur. At radio 
frequencies the phase separation between the normal main pulse and interpulse is 0.52106(3) (Kinkhabwala \& Thorsett 2000), 
while the separation between the locations of the giant radio pulses amounts 0.5264(6). The latter value is more consistent with the pulse separation at X-rays. Pulsed X-ray emission was detected up to $\sim 20$ keV and its spectrum could 
satisfactorily be described by a power-law with index $1.14\pm0.07$, consistent with the MECS derived value.

At soft and high-energy $\gamma$-rays only upper limits have been reported (CGRO OSSE: Schroeder et al. (1995)
; CGRO EGRET: Fierro et al. (1995)).
  
\section{Discussion \& conclusion}

\begin{figure}[t]
  \centerline{\psfig{figure=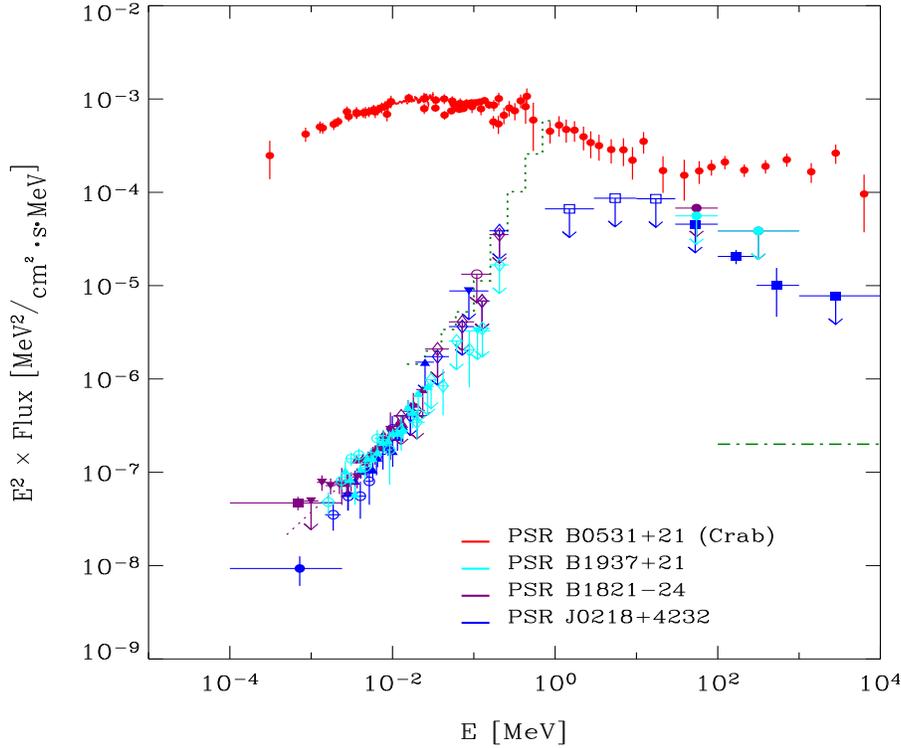,bbllx=85pt,bblly=200pt,bburx=500pt,bbury=615pt,clip=,height=10cm,width=12cm,angle=0}}
  \caption{Pulsed high-energy spectra of Class-II MSPs - PSR J1821-24, PSR J0218+4232
                 and PSR B1937+21 - in a $\nu F_{\nu}$ representation combining spectral information
                 obtained by different X-ray and $\gamma$-ray instruments. The pulsed high-energy spectrum
                 of the Crab pulsar is shown for reference. The $3\sigma$ sensitivity flux limits for 
                 the IBIS ISGRI detector aboard INTEGRAL assuming a 1 Ms observation are plotted as a dotted line.
                 The dash-dotted line between 100-10000 MeV is the GLAST $3\sigma$ sensitivity limit.
                 Notice the hard nature of the spectra at X-rays. The spectra soften dramatically at 
                 higher energies. The photon index for the $\gamma$-ray spectrum of PSR J0218+4232 is
                 about 2.6. The overall spectral shape is reminiscent to that of unidentified high-energy
                 $\gamma$-ray sources.\label{he_spectra}}
\end{figure}
The Class-II MSPs discussed in this review exhibit (very) hard pulsed spectra at X-ray energies up to $\sim$ 20 keV with photon 
indices in the range 1 -- 1.2. Their spectra must soften drastically in the hard X-ray or soft $\gamma$-ray range 
in order to be consistent with the
flux upper limits for PSR B1821-24 and PSR B1937+21 and the flux measurements of PSR J0218+4232 at $\gamma$-ray
energies (see Fig. 6). Such kind of spectra is reminiscent to that of unidentified high-energy $\gamma$-ray 
sources (UGS; Grenier et al. (1997), Gehrels et al. (2000)) and therefore Class-II MSPs could explain at least part 
of the UGSs. At high-energy $\gamma$-rays AGILE and in particular GLAST can contribute significantly to the
study of the high-energy properties of Class-II MSPs. Notably, the pulsed nature of the $\gamma$-ray
emission of PSR J0218+4232 can be studied in more detail. At soft $\gamma$-ray energies long observations ($>$ 1 Ms)
with INTEGRAL IBIS ISGRI can yield important spectral information beyond 20 keV.

Models attempting to describe the high-energy emission from spin-down powered pulsars can crudely be divided 
in two classes: polar cap models (e.g. Harding et al. (1981), Daugherty et al. (1996)) and outer-gap models 
(e.g. Cheng et al. (1986,2000), Romani et al. (1996)). In polar cap models the high-energy radiation is produced close to the polar
cap of the pulsar, while in outer gap models the high-energy radiation is generated in vacuum gaps near the last
close field lines between the null charge surface and the light cylinder. At X-ray energies both model types predict
a photon spectral index of $\sim 1.5$ (Rudak et al. (1999); Zhang et al. (2003)) which is too soft compared with the
measured indices of the Class-II MSPs. Also, in the $\gamma$-ray regime such spectral discrepancies exist. This underlines the
inability of the both classes of theoretical models to reproduce the emerging high-energy spectra for Class-II MSPs properly.
In Kuiper et al. (2000)  (see also Cognard et al. 1996, Saito et al.
1997, and Kuiper et al. 1998) it was noted that the three Class II MSPs
together with the Crab pulsar and its twin in the LMC PSR B0540-69 are
among the top six pulsars with the highest magnetic field strengths near
the light cylinder (the sixth one is MSP PSR B1957+20). It was argued
that the pulsed high-energy non-thermal emission of these pulsars
originates quite likely in outer gaps near the light cylinder, based on
similarities with the Crab pulsar. There is now additional supporting
evidence in favour of an outer gap scenario for the production of the 
non-thermal emission: Giant/large amplitude radio pulses - a phenomenon 
likely related to the physical conditions near the light cylinder (Romani \& Johnson 2001) -
have now been detected for all Class II MSPs (PSR B1937+21, Cognard et
al. 1996; PSR B1821-24, Romani \& Johnson 2001; PSR J0218+4232, Joshi et
al. 2003) as well as for the Crab (e.g. Lundgren et al. 1995) and very
recently PSR B0540-69 (Johnston \& Romani 2003). In fact, phase
coincidences between radio giant and X-ray pulses have been reported by
Romani and Johnson for PSR B1821-24 and by Cusumano et al. (2003) for
PSR B1937+21.
Future radio and high-energy observations of this intriguing class of MSPs are of great importance to shed light 
on the origin and nature of the emerging electro-magnetic radiation.  



\end{document}